\documentclass[preprint,aps,showpacs]{revtex4}

\usepackage{graphicx}
\usepackage{amsfonts}
\usepackage{amsmath}

\begin{document}
\title{Testing for Markovian Character and
Modeling of Intermittency in Solar Wind Turbulence}
\author{Marek Strumik}
\email{maro@cbk.waw.pl}
\affiliation{Space Research Centre, Polish Academy of Sciences,
Bartycka 18 A, 00-716 Warsaw, Poland}
\author{Wies{\l}aw M. Macek}
\affiliation{Faculty of Mathematics and Natural Sciences. College of Sciences,
Cardinal Stefan Wyszy\'{n}ski University,
Dewajtis 5, 01-815 Warsaw;\\
Space Research Centre, Polish Academy of Sciences,
Bartycka 18 A, 00-716 Warsaw, Poland}
\date{\today}
\begin{abstract}
We present results of statistical analysis of solar wind turbulence
using an approach based on the theory of Markov processes. It is shown that
the Chapman-Kolmogorov equation is approximately satisfied for the turbulent
cascade. We evaluate the first two Kramers-Moyal coefficients from
experimental data and show that the solution of the resulting
Fokker-Planck equation agrees well with experimental probability
distributions. Our results suggest the presence of a local transfer mechanism
for magnetic field fluctuations in solar wind turbulence.
\end{abstract}
\maketitle

\section{Introduction}
Irregular dynamics of the solar wind plasma exhibits many similarities to
fully developed hydrodynamic turbulence. Numerous in situ measurements
of temporal variability of parameters of the plasma have shown
that their spectral distributions
usually have power-law character \cite{MatGol82,Goletal95,TuMar95,GolRob99}.
Investigations of the fluctuations have also revealed their non-Gaussian
probability distributions at small scales, which is
commonly attributed to intermittency phenomenon
\cite{MarTu94,Soretal99,Soretal01,Bur01}. In fact, the solar wind provides a
unique laboratory for studying high-Reynolds-number magnetohydrodynamic
turbulence (see, e.g., Refs. \cite{Goletal95,BruCar05} for review).

One of the main problems in the studies of incompressible hydrodynamical
turbulence is explaining the statistics of velocity fluctuations on
different length scales. In magnetohydrodynamic turbulence this problem
concerns in general also magnetic field and density fluctuations.
Conventionally, in investigations of a turbulent cascade, statistical
properties of fluctuations $\delta u_\tau(t)=u(t+\tau)-u(t)$
of a physical quantity $u(t)$ are examined, where $\tau$ is temporal 
(or spatial) scale.
The fluctuations are studied by examining their probability distribution
functions (p.d.f.)
$P(\delta u_\tau(t))$ or $n$-order moments $\langle \delta u_\tau(t)^n \rangle$
of the distributions, called also structure functions.
Often, if the root-mean-square of velocity fluctuations is small
as compared to the mean velocity of the flow, one can use the Taylor hypothesis,
interpreting the temporal variation $\delta u_\tau$ at a fixed position
as a spatial variation $\delta u_l$, where $l$ is a spatial scale
corresponding to the temporal scale $\tau$.
In an intermittent turbulent cascade, the p.d.f. of the fluctuations is
non-Gaussian at small scales. When we go to larger scales, the shape of the
p.d.f. changes, and finally there is a scale $\tau_G$, such that for
$\tau>\tau_G$ the p.d.f. is close to Gaussian distribution \cite{Fri95,Bis03}.

A number of models for the scaling exponents and scaling of the
probability distributions of the fluctuations have been proposed.
Many papers have been also devoted to experimental verification of 
the proposed models (see, e.g., Refs. \cite{Fri95,Bis03,BruCar05} for
review). Recently, a great deal of attention has been devoted to
investigations of the fluctuations in hydrodynamic turbulence
from the point of view of the Markov processes theory (see, e.g.,
Refs. \cite{PedNov94,FriPei97a,FriPei97b,DavTab99,Frietal00,Renetal01}).
In particular, results of the verification of the validity of the
Chapman-Kolmogorov equation as well as estimations of the Kramers-Moyal
coefficients from experimental data suggest that the Markov
processes approach may be appropriate to the description and
modeling of the turbulent cascade \cite{FriPei97a,FriPei97b,Renetal01}.
The estimations of the Kramers-Moyal coefficients allow to
determine the form of the Fokker-Planck equation governing the evolution
of the probability distribution with scale for the fluctuations.
A model based on a Fokker-Planck equation has been recently
proposed for solar wind turbulence, but for fluctuations of quantities
that exhibit self-similar scaling \cite{Hnaetal03}.
In the present paper, the Markov processes approach for
the first time has been applied to analysis of intermittent
solar wind turbulence. 
This approach seems to provide a contact point between pure statistics
and dynamical systems approach to turbulence.

\section{Data Set}
In the plasma flow expanding from the Sun into the interplanetary space
we can distinguish several forms, in particular the slow
($< 450$ km/s) and
fast ($> 600$ km/s) solar wind (see, e.g., Ref. \cite{Sch06} and references
therein). At the solar minimum the two forms are usually well separated,
the fast wind is more homogeneous and incompressible in comparison with
the slow wind. Our goal here is to study properties of the turbulent
cascade, therefore we try to exclude effects associated with nonstationary
driving and spatial inhomogeneity of the turbulence. For this reason,
in this paper we have chosen for analysis the fast solar wind flowing
from non-active high-latitude regions in the solar corona at the solar
minimum. This data set represents dynamics of fast solar wind
free of dynamical interaction with slow wind, as possibly the most
homogeneous and probably also most stationary case. Therefore effects
associated with nonstationary driving should be eliminated
here to a large extent, and we should observe a state possibly
closest to freely decaying turbulence, which seems to be the most
appropriate case to study the turbulent cascade.
Since we would like to examine fluctuations in a wide range of scales,
including small scales, we focus here on magnetic field fluctuations, which
are available at much better time resolution in comparison with measurements
of plasma parameters (e.g., bulk velocity or density of the plasma).
However, we are aware of the importance of detailed analysis of other types
of the solar wind, as well as other bulk plasma parameters, and we are
going to carry out such studies in a future.

The data set analyzed here consists of about $1.3 \times 10^7$ measurements
of the radial component $B_R$ of the magnetic field obtained by the Ulysses
spacecraft from 70:1996 to 230:1996 (day of year:year) at time resolution
of one second (see Ref. \cite{Baletal92} for the description of the
experimental setup).
The measurements have been obtained at heliospheric
latitudes from 29 to 44 degree and at radial distance from the Sun
from 3.5 to 4.2 AU. Small gaps (up to three missing points) in the data
set have been filled using linear interpolation. 
Further in this paper we consider fluctuations of the radial component of
the magnetic field defined as $b_\tau(t)=B_R(t+\tau)-B_R(t)$. 
Presenting results of our analysis, we
do not recast the fluctuations into the space domain via the
Taylor hypothesis, i.e., we consistently use here temporal scales.
However, since we analyze highly supersonic and super-Alfv\'{e}nic
flow (mean velocity $U \approx 744$ km/s in the reference system moving
with the measuring instrument), the temporal scales physically should 
be interpreted rather as spatial scales. Assuming that the Taylor hypothesis
is satisfied here, one can easily transform the temporal scale $\tau$ to
the spatial scale $l$ using the relationship $l=U\tau$ \cite{Fri95}.
However, since Ulysses
spacecraft provides one-point measurements of the magnetic field, in general
it is not possible to distinguish between temporal and spatial variations in
this case.

\begin{figure}[!thb]
\begin{center}
\includegraphics[scale=0.7]{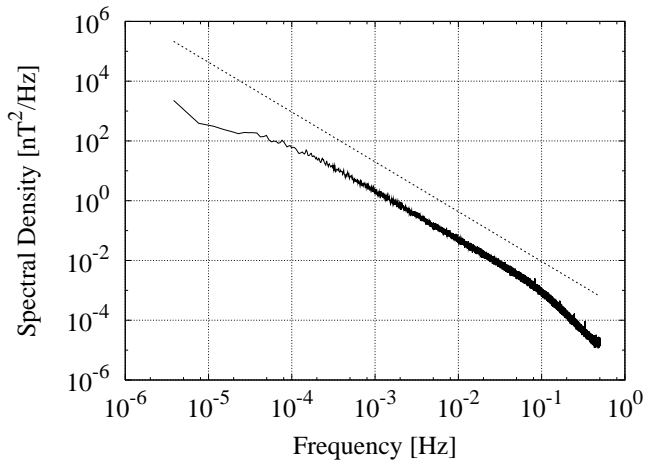}
\caption{Power spectrum of the radial component of the solar wind magnetic
field (solid line). The dashed line shows the spectrum of the type
$E(f) \propto f^{-5/3}$ for comparison.}
\label{fig:spectr}
\end{center}
\end{figure}

In Fig. \ref{fig:spectr} we show the power spectrum of the radial component
of the magnetic field. As one can see, the power spectrum
has a power-law character with spectral exponent very close to -5/3 in the
inertial range identified here as stretching approximately from 0.0002 to
0.075 Hz.

\section{Markov processes approach}
In the case of a Markov process, by definition the following condition
must be satisfied
\begin{equation}
P(b_{\tau_1},\tau_1|b_{\tau_2},\tau_2; \ldots ;b_{\tau_N},\tau_N)=
P(b_{\tau_1},\tau_1|b_{\tau_2},\tau_2),
\end{equation}
thus the $N$-point joint p.d.f.
$P(b_{\tau_1},\tau_1;b_{\tau_2},\tau_2; \ldots ;b_{\tau_N},\tau_N)$
is determined by the product of conditional probabilities
$P(b_{\tau_{i-1}},\tau_{i-1}|b_{\tau_{i}},\tau_{i})$, where
$\tau_{i-1}<\tau_{i}$. For a finite set of experimental data, the
Markov property can be verified by comparison of a conditional
p.d.f. $P_{\mathrm{E}}(b_{\tau_1},\tau_1|b_{\tau_2},\tau_2)$ evaluated
directly from data with the p.d.f. computed using the Chapman-Kolmogorov
equation
\begin{equation}
P(b_{\tau_1},\tau_1|b_{\tau_2},\tau_2)=\int_{-\infty}^{\infty} P(b_{\tau_1},\tau_1|b_{\tau'},\tau')
P(b_{\tau'},\tau'|b_{\tau_2},\tau_2) \mathrm{d}b_{\tau'},
\label{eq:chk}
\end{equation}
where $\tau_1<\tau'<\tau_2$. Eq. (\ref{eq:chk}) is
a necessary condition for a stochastic process to be Markovian.
The Chapman-Kolmogorov equation can be written in a differential
form using the so-called Kramers-Moyal expansion
\begin{equation}
-\tau\frac{\partial P(b_\tau,\tau|b_{\tau_0},\tau_0)}{\partial \tau} = \sum_{k=1}^{\infty}
\left( -\frac{\partial}{\partial b_\tau} \right)^k
D^{(k)}(b_\tau,\tau) P(b_{\tau},\tau|b_{\tau_0},\tau_0).
\label{eq:kme}
\end{equation}
Kramers-Moyal coefficients $D^{(k)}(b_{\tau},\tau)$ can be evaluated as 
the limit $\Delta \tau \rightarrow 0$ of the conditional moments
$M^{(k)}(b_{\tau},\tau,\Delta \tau)$, namely
\begin{equation}
D^{(k)}(b_{\tau},\tau)=\lim_{\Delta \tau \rightarrow 0}M^{(k)}(b_{\tau},\tau,\Delta \tau)
\label{eq:kmcoeff}
\end{equation}
and
\begin{equation}
M^{(k)}(b_{\tau},\tau,\Delta \tau)=\frac{\tau}{k! ~ \Delta \tau} \int_{-\infty}^{\infty}
(b_{\tau'}-b_\tau)^k P(b_{\tau'},\tau'|b_{\tau},\tau) \mathrm{d}b_{\tau'},
\label{eq:condmoms}
\end{equation}
where $\Delta \tau =\tau -\tau'$.
If $D^{(4)}(b_{\tau},\tau)=0$ then according to the Pawula theorem:
$D^{(k)}(b_{\tau},\tau)=0$ for $k \ge 3$ \cite{Ris89}.
In this case, starting from Eq. (\ref{eq:kme}) we arrive at the
Fokker-Planck equation
\begin{equation}
-\tau\frac{\partial P(b_{\tau},\tau)}{\partial \tau}=
\left( -\frac{\partial D^{(1)}(b_{\tau},\tau)}{\partial b_\tau}+
\frac{\partial^2 D^{(2)}(b_{\tau},\tau)}{\partial b_\tau^2} \right) P(b_{\tau},\tau),
\label{eq:fp}
\end{equation}
which determines the evolution of the probability distribution function
of a stochastic process generated by the Langevin equation
(Ito definition)
\begin{equation}
-\tau\frac{\mathrm{d} b_{\tau}}{\mathrm{d} \tau}=
D^{(1)}(b_{\tau},\tau) + \sqrt{D^{(2)}(b_{\tau},\tau)} ~ \Gamma(\tau),
\end{equation}
where $\Gamma(\tau)$ is the delta-correlated Gaussian noise.
In comparison with definition used in Ref. \cite{Ris89}, the Kramers-Moyal
coefficients given here are multiplied by $\tau$, which is equivalent to
a logarithmic length scale \cite{Renetal01}.

If Eq. (\ref{eq:chk}) is satisfied, then the transition
probability from scale $\tau_2$ to $\tau_1$
can be divided into transitions from $\tau_2$ to $\tau'$ and
then from $\tau'$ to $\tau_1$. Therefore,
in the case of a turbulent cascade, fulfillment of the Chapman-Kolmogorov
equation for all triplets $\tau_1<\tau'<\tau_2$ in the inertial range
suggests the presence of a local transfer mechanism in the
cascade.

\section{Results}
\begin{figure}[!thb]
\begin{center}
\includegraphics[scale=0.7]{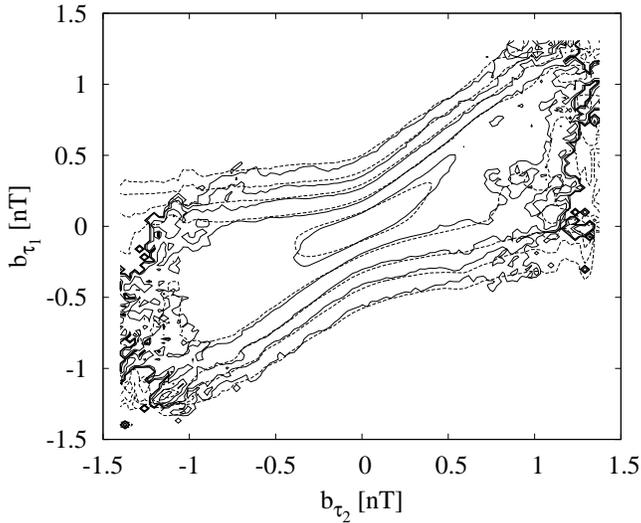}
\caption{Contour plots illustrating verification of the Chapman-Kolmogorov
equation for $\tau_1=750$, $\tau'=1000$, and $\tau_2=1250$
seconds. Solid lines represent the conditional p.d.f.
$P_{\mathrm{E}}(b_{\tau_1},\tau_1|b_{\tau_2},\tau_2)$ evaluated directly from
data, whereas dashed lines show the conditional p.d.f. 
$P(b_{\tau_1},\tau_1|b_{\tau_2},\tau_2)$ computed using Eq.
(\ref{eq:chk}).
The subsequent isolines correspond to the following levels of
the p.d.f.: 2.0, 0.7, 0.2, 0.07, 0.02 (from the middle of the plot).}
\label{fig:chk1}
\end{center}
\end{figure}

\begin{figure}[!thb]
\begin{center}
\includegraphics[scale=0.7]{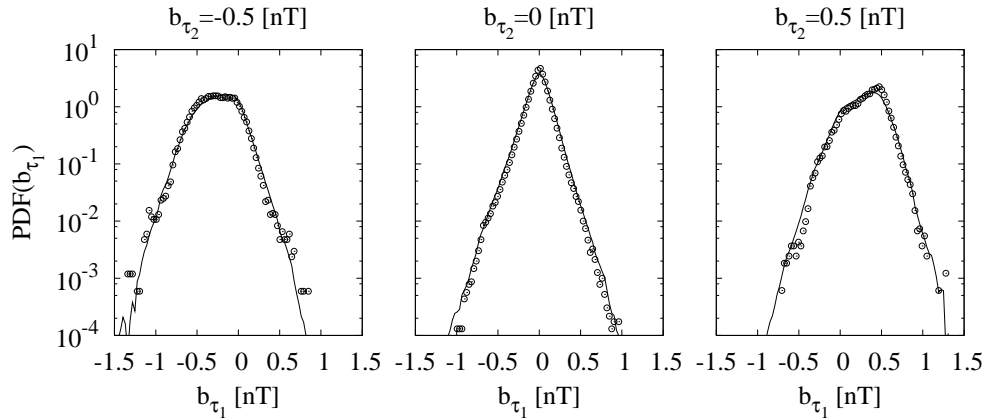}
\caption{Verification of the Chapman-Kolmogorov equation
(\ref{eq:chk}) for $\tau_1=750$, $\tau'=1000$, and $\tau_2=1250$ seconds.
Comparisons of cuts through $P_{\mathrm{E}}(b_{\tau_1},\tau_1|b_{\tau_2},\tau_2)$
(points) and $P(b_{\tau_1},\tau_1|b_{\tau_2},\tau_2)$ (lines) from
Fig. \ref{fig:chk1} are shown
for fixed values of $b_{\tau_2}$, namely $b_{\tau_2}=-0.5$ nT (on the left),
$b_{\tau_2}=0$ nT (in the middle), and $b_{\tau_2}=0.5$ nT (on the right).}
\label{fig:chk2}
\end{center}
\end{figure}

In Fig. \ref{fig:chk1} we show superposed contour plots of the
conditional p.d.f. estimated directly
from data and the p.d.f. computed using Eq. (\ref{eq:chk})
for $\tau_1=750$, $\tau'=1000$, and $\tau_2=1250$ seconds. One can see that
corresponding contour lines for the two probability distributions are very
close to each other. This indicates that the Chapman-Kolmogorov equation is
(at least approximately) satisfied for the range of scales from $\tau_1=750$
to $\tau_2=1250$ seconds. In Fig. \ref{fig:chk2} we show the
cuts through the conditional probability distributions for fixed
values of $b_{\tau_2}$. As one can see, points representing cuts through
$P_{\mathrm{E}}(b_{\tau_1},\tau_1|b_{\tau_2},\tau_2)$ fit well to the lines 
representing cuts through $P(b_{\tau_1},\tau_1|b_{\tau_2},\tau_2)$.
Repeating such a comparison for different
triplets $\tau_1,\tau',\tau_2$ we have checked that Eq. (\ref{eq:chk})
is well satisfied in the inertial range (for scales from about 50 to
5000 seconds). For larger scales, outside the inertial range,
the larger is the scale, a worse agreement we observe between
experimental p.d.f. and that computed using Eq. (\ref{eq:chk}).
Nevertheless, Eq. (\ref{eq:chk}) seems to be fulfilled up to the scale
of about 24 hours. Therefore, the necessary condition for
Markov processes is satisfied here in the entire range of scales
available for our computations, unlike it is in the case of
hydrodynamic turbulence as reported in Ref. \cite{Renetal01}, where
the cascade is not Markovian for small scales, below the Taylor length
scale.

\begin{figure}[!thb]
\begin{center}
\includegraphics[scale=0.7]{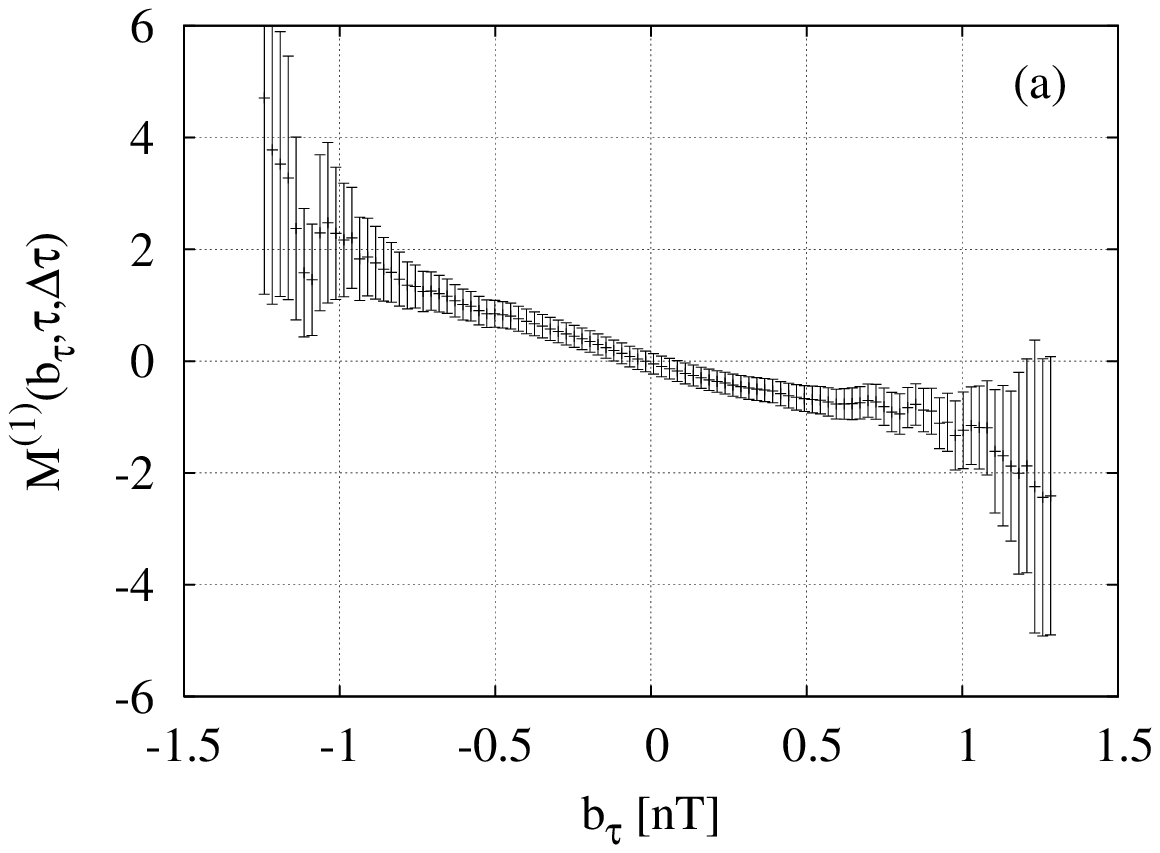}
\includegraphics[scale=0.7]{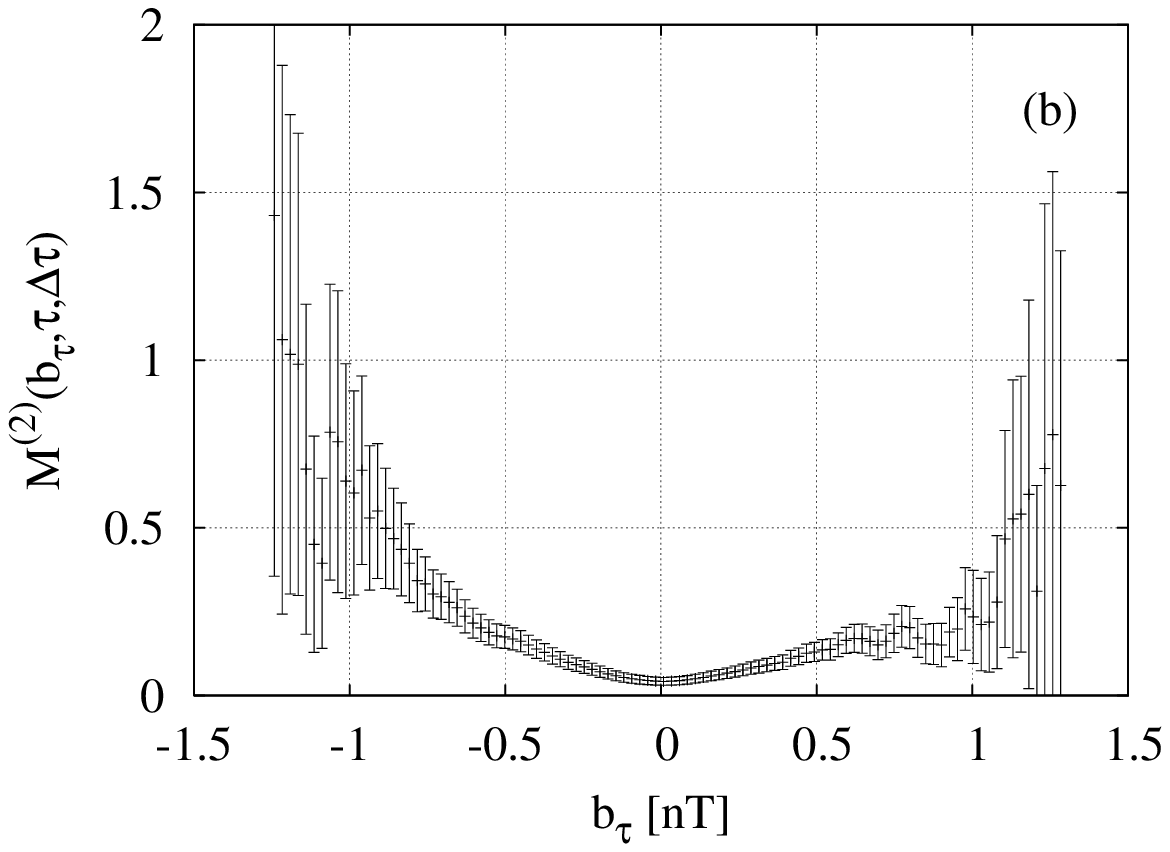}
\includegraphics[scale=0.7]{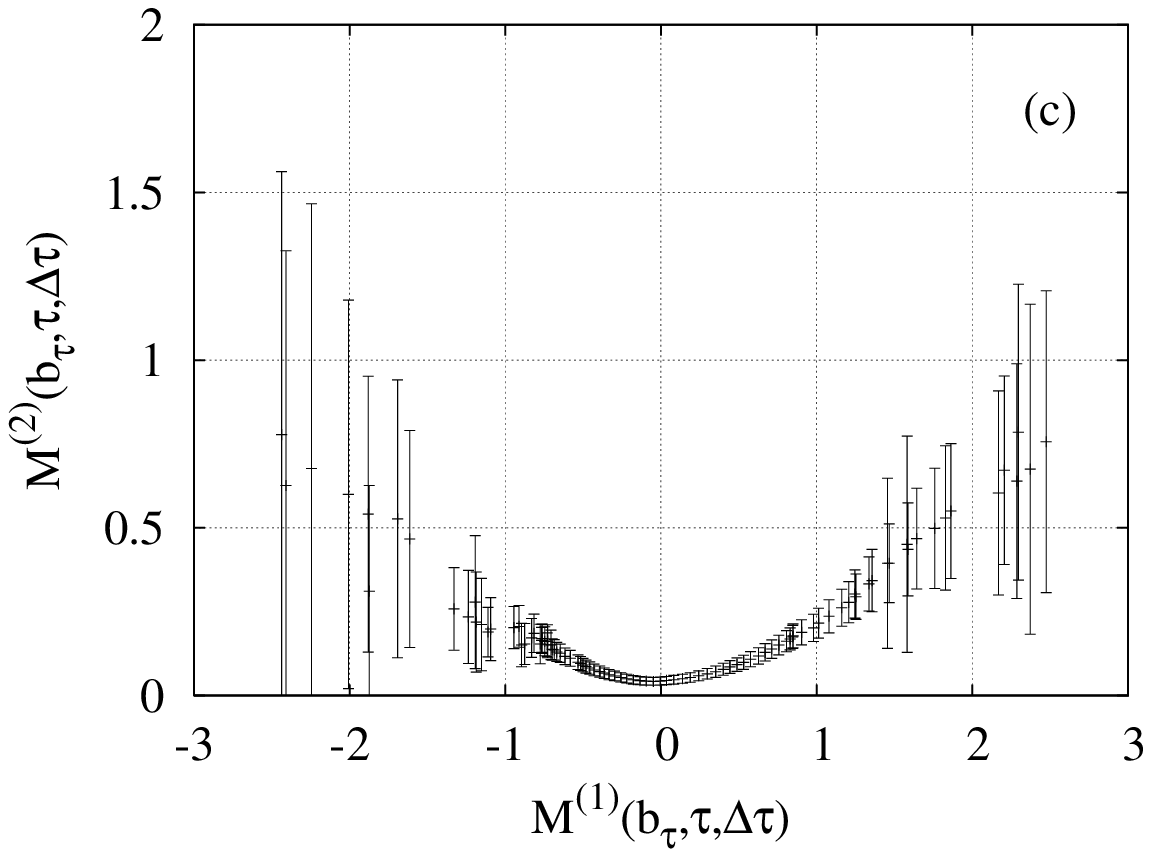}
\caption{The dependence of the coefficients (a)
$M^{(1)}(b_{\tau},\tau,\Delta \tau)$ and (b)
$M^{(2)}(b_{\tau},\tau,\Delta \tau)$
on $b_{\tau}$ for $\tau=1000$ and $\Delta \tau = 100$ seconds.
In panel (c) we show the dependence of 
$M^{(2)}(b_{\tau},\tau,\Delta \tau)$ on $M^{(1)}(b_{\tau},\tau,\Delta \tau)$.}
\label{fig:kmcoeff}
\end{center}
\end{figure}

\begin{figure}[!thb]
\begin{center}
\includegraphics[scale=0.7]{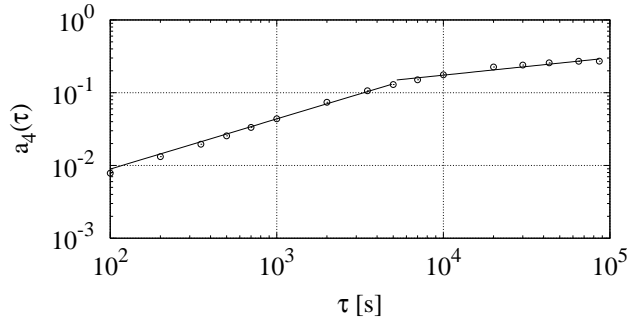}
\caption{Dependence of the parameter $a_4$ on scale $\tau$ (see Eq.
(\ref{eq:d2fit})).}
\label{fig:a4tau}
\end{center}
\end{figure}

We have computed the coefficients $M^{(k)}(b_{\tau},\tau,\Delta \tau)$
using the definition of Eq. (\ref{eq:condmoms}). In Figs. \ref{fig:kmcoeff}(a)
and \ref{fig:kmcoeff}(b)
we present examples of the dependence of the coefficients
$M^{(1)}(b_{\tau},\tau,\Delta \tau)$ and $M^{(2)}(b_{\tau},\tau,\Delta \tau)$
on $b_{\tau}$ for $\tau=1000$ and $\Delta \tau = 100$ seconds.
In Fig. \ref{fig:kmcoeff}(c) we show the dependence of 
$M^{(2)}(b_{\tau},\tau,\Delta \tau)$ on $M^{(1)}(b_{\tau},\tau,\Delta \tau)$,
which have a more regular and symmetric form in comparison with
the dependence of $M^{(2)}(b_{\tau},\tau,\Delta \tau)$ on $b_{\tau}$.
We propose the following approximations:
$M^{(1)}(b_{\tau},\tau,\Delta \tau)=A_1(\tau,\Delta \tau) b_\tau +
A_2(\tau,\Delta \tau) b^3_\tau + A_3(\tau,\Delta \tau) b^5_\tau$
and
$M^{(2)}(b_{\tau},\tau,\Delta \tau)=A_4(\tau,\Delta \tau) +
A_5(\tau,\Delta \tau) [M^{(1)}(b_{\tau},\tau,\Delta \tau)]^2$
as describing properly the experimental relationships shown
in Figs. \ref{fig:kmcoeff}(a) and \ref{fig:kmcoeff}(c),
correspondingly. Based on the
approximations, we can fit the parameters $A_i(\tau,\Delta \tau)$ for
a fixed $\tau$ and changing $\Delta \tau$, and finally compute the limits 
$a_i(\tau)=\lim_{\Delta \tau \rightarrow 0}A_i(\tau,\Delta \tau)$
for $i=1\ldots5$ (e.g., by a linear extrapolation toward $\Delta \tau=0$)
obtaining the following approximations:
\begin{equation}
D^{(1)}(b_{\tau},\tau)=a_1(\tau) b_\tau + a_2(\tau) b^3_\tau +
a_3(\tau) b^5_\tau
\label{eq:d1fit}
\end{equation}
and
\begin{equation}
D^{(2)}(b_{\tau},\tau)=a_4(\tau)+a_5(\tau) [D^{(1)}(b_{\tau},\tau)]^2.
\label{eq:d2fit}
\end{equation}
Repeating the entire procedure for changing $\tau$ we can also
estimate the dependence of the coefficients $a_i$ on $\tau$.
Applying the algorithm, we have obtained the following results
for the inertial range ($\tau \le 5000$ seconds) $
a_1=-3.6\tau^{-0.08}, a_2=3.5 \exp(-0.0001\tau), a_3=-13.6\tau^{-0.2},
a_4=0.00035\tau^{0.7}, a_5=1.2\tau^{-0.3}$,
and outside the inertial range ($\tau>5000$ seconds)
$a_1=-0.5\tau^{0.16}, a_2=2, a_3=-2.3,
a_4=0.016\tau^{0.26}, a_5=1.75\tau^{-0.36}$.
As an illustration, in Fig. \ref{fig:a4tau} we show the dependence of the 
parameter $a_4$ on $\tau$. One can notice a change in the
dependence for $\tau \approx 5000$ seconds, i.e., at the end of
inertial range.

\begin{figure}[!thb]
\begin{center}
\includegraphics[scale=0.7]{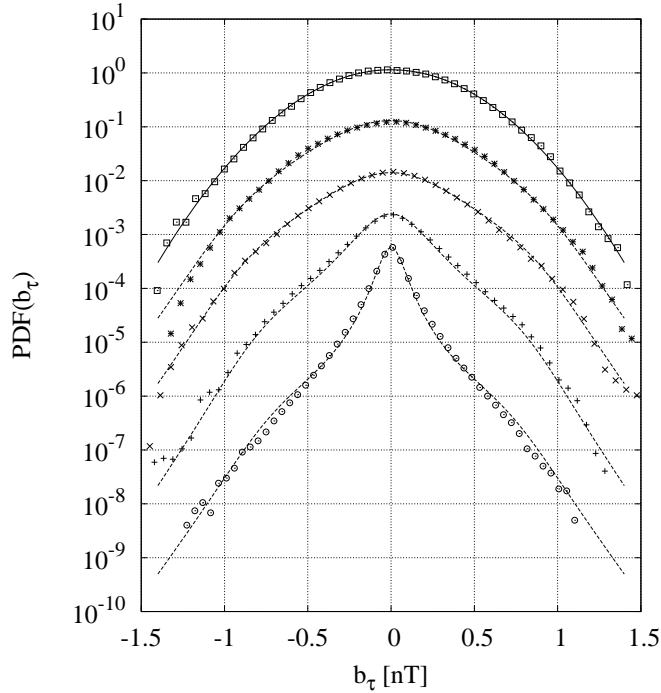}
\caption{Experimental probability distributions (points) and solution of 
the Fokker-Planck equation (dashed lines) for initial condition (solid line)
obtained by approximation of the experimental p.d.f. by Gaussian distribution
for $\tau=86400$ seconds. We show the comparison of the experimental p.d.f.
and solution of Eq. (\ref{eq:fp}) for $\tau$ equal to 86400, 30000, 5000,
1000, 100 seconds (from the top). Probability distributions for different
scales have been shifted in the vertical direction for clarity of
presentation.}
\label{fig:pdfcomp}
\end{center}
\end{figure}

Parameterization of $D^{(1)}(b_{\tau},\tau)$ and $D^{(2)}(b_{\tau},\tau)$
(shown in Eqs. (\ref{eq:d1fit}) and (\ref{eq:d2fit}) correspondingly)
with experimentally fitted parameters $a_i(\tau)$
allows us to solve numerically Eq. (\ref{eq:fp})
with initial condition taken from parameterization of the experimental
p.d.f. at a large scale $\tau_G$, where the probability distribution of 
fluctuations is approximately Gaussian. Therefore we can compute
numerically p.d.f. at scales $\tau<\tau_G$ and compare it to the
experimental p.d.f., which allows us to verify directly our results.
Such a comparison is shown in Fig. \ref{fig:pdfcomp} for 
$\tau_G=86400$ seconds. As one can see
there is a good agreement between experimental probability distributions and
those computed from the Fokker-Planck equation.

\section{Discussion}
We have shown that the Markov processes approach can be
applied to the description of the turbulent cascade in fast solar wind
turbulence. The Chapman-Kolmogorov equation is approximately satisfied in
the inertial range, as well as for larger scales up to $\tau=86400$ seconds.
Numerical solution of the Fokker-Planck equation agrees well with
experimental probability distributions obtained directly from data
in the range of $\tau$ from 100 to 86400 seconds. Therefore, we can conclude
that for intermittent solar wind turbulence, the Markov processes approach 
can provide a mathematical formalism capable of explaining specific
evolution of the shape of the probability distribution with scale changing
from energy containing scale down to dissipation scale. Since the formalism
describes properly evolution of the probability distribution with scale,
obviously this should also work for structure functions, which are defined as 
appropriate moments of the probability distributions. Admittedly, direct
analytical derivation of scaling properties of the structure functions
can be difficult, but we expect that such studies can be done numerically.

Every Markov process must satisfy the Chapman-Kolmogorov equation (\ref{eq:chk}),
which express the condition that the probability density of transition
from scale $\tau_2$ to $\tau_1$, can be subdivided into smaller steps,
that is transition from scale $\tau_2$ to $\tau'$, and then
from scale $\tau'$ to $\tau_1$. Therefore, in the case of a turbulent
cascade, fulfillment of the Chapman-Kolmogorov equation can be interpreted
as the presence of a local transfer mechanism in wave vector space.
The question of locality of the energy transfer is of some interest,
e.g., in the studies of dynamo mechanism to generate magnetic fields in
astrophysical objects, where in helical MHD turbulence, nonlocal
processes of generation of large-scale fields by small-scale helicities
are studied in details (see, e.g., Sec. 6.2.1 of \cite{Bis03}).
The question is also important for modeling MHD flows and numerical
simulations, e.g.,
in large-eddy simulations, where low-pass filtering with respect to a cutoff
wave number requires some assumptions concerning the transfer of energy
around the cutoff wavenumber.
Local and nonlocal transfer mechanisms can be distinguished in theoretical
studies of turbulence via shell models or numerical simulations
(see, e.g., \cite{Debetal05,Aleetal05,Minetal05}), but it is
very difficult to study the property of turbulence using experimental data.
The Markov processes approach seems to provide such a method.
Namely, analyzing a time
series from a turbulent flow we should be able to identify the character
of the dominating transfer mechanism for a given quantity or between
different quantities, i.e., we should be able to answer the question
whether the mechanism is local or nonlocal.

Since our results suggest rather the Markovian character of the turbulent
cascade in the solar wind, it indicates that the local transfer mechanism 
dominates in solar wind turbulence. Therefore dominating transfer of
magnetic field fluctuations has similar character as in the case of
Kolmogorov phenomenology describing turbulence in neutral fluids,
where according to the picture of Richardson
cascade, the energy transfer has local character
in the wave vector space, i.e., the energy at a scale $l$ is transfered
mainly to comparable scales \cite{Fri95}. This result is somewhat
surprising, because we analyze here magnetohydrodynamic turbulence,
which is rather magnetic field dominated case. Therefore, according to the
classical Iroshnikov-Kraichnan picture, due to the Alfv\'{e}n effect, we should
expect nonlocal influence of large-scale magnetic field on small-scale
turbulent eddies, and so also nonlocal interactions between modes \cite{Bis03}.
However, results of recent numerical simulations suggest
that local transfer mechanisms are present in MHD turbulence
\cite{Debetal05,Aleetal05,Minetal05}.
Our paper provides experimental results confirming this observation for
magnetic-to-magnetic field transfer.

\acknowledgments
This work has been supported by the Polish Ministry of Science and Higher
Education through Grant No. \mbox{N N202 4127 33}.


\end{document}